\documentclass{article}

\usepackage{PRIMEarxiv}

\usepackage[utf8]{inputenc} 
\usepackage[T1]{fontenc}    
\usepackage{url}            
\usepackage{booktabs}       
\usepackage{amsfonts}       
\usepackage{nicefrac}       
\usepackage{microtype}      
\usepackage{lipsum}
\usepackage{fancyhdr}       
\usepackage{graphicx}       
\graphicspath{{media/}}     

\usepackage{amsmath}
\usepackage{mathtools}
\usepackage{bm}
\usepackage{subcaption}
\usepackage{booktabs}
\usepackage{multirow}
\usepackage{setspace}
\usepackage{caption}
\usepackage[flushleft]{threeparttable}

\title{Identifying Hidden Visits from Sparse Call Detail Record Data
}

\author{
  Zhan Zhao\thanks{Corresponding Author} \\
  Department of Urban Planning and Design \\
  The University of Hong Kong \\
  Hong Kong SAR, China\\
  \texttt{zhanzhao@hku.hk} \\
   \And
  Haris N. Koutsopoulos \\
  Department of Civil and Environmental Engineering \\
  Northeastern University \\
  Bonston, MA, United States\\
  \texttt{h.koutsopoulos@northeastern.edu} \\
   \And
  Jinhua Zhao \\
  Department of Urban Studies and Planning \\
  Massachusetts Institute of Technology \\
  Cambridge, MA, United States\\
  \texttt{jinhua@mit.edu} \\
}

\begin{document}
\maketitle

\begin{abstract}
Despite a large body of literature on trip inference using call detail record (CDR) data, a fundamental understanding of their limitations is lacking. In particular, because of the sparse nature of CDR data, users may travel to a location without being revealed in the data, which we refer to as a \textit{hidden visit}. The existence of hidden visits hinders our ability to extract reliable information about human mobility and travel behavior from CDR data. In this study, we propose a data fusion approach to obtain labeled data for statistical inference of hidden visits. In the absence of complementary data, this can be accomplished by extracting labeled observations from more granular cellular data access records, and extracting features from voice call and text messaging records. The proposed approach is demonstrated using a real-world CDR dataset of 3 million users from a large Chinese city. Logistic regression, support vector machine, random forest, and gradient boosting are used to infer whether a hidden visit exists during a displacement observed from CDR data. The test results show significant improvement over the naive no-hidden-visit rule, which is an implicit assumption adopted by most existing studies. Based on the proposed model, we estimate that over 10\% of the displacements extracted from CDR data involve hidden visits. The proposed data fusion method offers a systematic statistical approach to inferring individual mobility patterns based on telecommunication records.
\end{abstract}

\keywords{Call Detail Record \and Individual mobility \and Trip inference \and Data fusion \and Hidden visit}

\section{Introduction}

Enabled by the increasing availability of large-scale datasets on human movements, human mobility has become an emerging field dedicated to extracting patterns that describe individual trajectories in time and space. In its essence, human movements are results of spatiotemporal choices (e.g., the decision to go somewhere at some time) made by individuals with diverse preferences and lifestyles. Trips reflect critical travel decisions, and thus are basic behavioral units of human mobility. A trip is defined as ``the travel required from an origin location to access a destination for the purpose of performing some activity'' \cite{mcnally_four-step_2007}. The ability to extract trips from large-scale spatiotemporal data sources is important for urban planning, transportation management, and location-based services.

One of the most commonly used data sources for human mobility studies is call detail record (CDR) data, which are collected by cellular service operators primarily for billing information collection and network management. CDR data are one type of event-driven mobile phone network data \cite{calabrese_urban_2014}. The generative events typically include incoming and outgoing voice calls, text messages (or Short Message Service, SMS), and, in some cases, cellular data usage (e.g., 3G/4G). In this study, we treat cellular data usage records as referred to in \cite{ranjan_are_2012}, as part of CDR data. Whenever a cellular transaction is made, the CDR database records its time and approximate location, in the form of the connected cell tower or antenna. Thus, CDR data offer the opportunity to capture spatiotemporal patterns of mobile phone users over time at a large scale. 

In recent years, CDR data have been used extensively to extract useful human mobility patterns and urban transportation information. The related studies cover diverse topics ranging from origin-destination (OD) estimation \cite{caceres_deriving_2007,calabrese_estimating_2011,mellegard_origin/destination-estimation_2011,wang_estimating_2013,iqbal_development_2014} and travel time estimation \cite{hasan_estimating_2017}, to meaningful place detection \cite{ahas_using_2010,isaacman_identifying_2011} and human activity discovery \cite{gonzalez_understanding_2008,phithakkitnukoon_activity-aware_2010,schneider_unravelling_2013,csaji_exploring_2013}. The majority of these studies depend on the ability to accurately extract trips from CDR data. However, unlike Global Positioning System (GPS) data (e.g., \cite{zhao_explaining_2015}), CDRs are recordings of people's telecommunication activities, which are not perfectly aligned with their travel behavior \cite{xu_uncovering_2018}. This raises the need to translate a series of telecommunication activities into a series of travel activities, which is not a straightforward task \cite{zhao_individual-level_2016}.

One critical limitation of CDR data for trip extraction is its sparsity. Phone usage tends to be sporadic in nature \cite{barabasi_origin_2005}. For most users, their mobile phone records are sparsely and irregularly distributed over time, resulting in periods when users may travel but have no phone records to reveal it in the CDR data. We call these time periods \textit{elapsed time intervals}, or ETIs. An ETI is defined as the period between two consecutive mobile phone records that is long enough for a user to potentially make a trip unobservable from the CDR data. When ETIs occur, the observed spatiotemporal traces of the user are likely incomplete, and the trip estimations based on such incomplete observations are prone to errors. For example, because of the sparsity issue, the fact that two locations are sequentially observed in CDR data does not mean that they are connected by a direct trip. They may not be an OD pair if the user makes an unobserved trip to another location between them. In other words, there may be a \textit{hidden visit}, which occurs when a user visits a place but has no CDR associated with it. By definition, hidden visits can only occur during ETIs. This issue has received limited attention in existing literature \cite{bayir_mobility_2010,chen_complete_2019}. Without properly considering hidden visits, the extracted OD pairs may be incorrect, the trip generation rate may be underestimated, and the spatiotemporal distribution of trips is likely to be skewed based on individual preferences of mobile phone usage \cite{zhao_understanding_2016}. This calls for methods that can infer the existence of hidden visits based on spatiotemporal context of the ETI as well as the individual characteristics of the user.

The objective of this study is to highlight the issue of hidden visits, and develop an approach to infer the existence of hidden visits during ETIs. Inferring something unobservable in the data, using the said data, is a challenging task. Typically, an unsupervised approach (e.g., \cite{chen_complete_2019}) is the only choice. However, the heterogeneity across different subsets of CDR data, e.g., voice call vs data access records, raises the opportunity to adopt a supervised approach based on data fusion for hidden visit inference. Specifically, for a subgroup of users with passively generated data activities, their data access records may be used to recover the portion of travel that is hidden from actively generated voice call records, which can then be used to train hidden visit inference models applied to general user population. In this paper, we focus on the problem of inferring whether hidden visits exist or not. The ability to identify hidden visits is important for ensuring the quality of the trip-level information extracted from CDR data. For example, if a hidden visit exists, the extracted OD pair should not be used for OD estimation. By identifying hidden visits during ETIs, we may distinguish OD pairs that are accurately inferred from those that are not. It is worth emphasizing that this paper focuses on identifying the existence of hidden visits, i.e., a binary problem, which is of great value for trip extraction by itself. It will provide a foundation for the inference of the exact time and location of the hidden visits, which is a relatively more challenging problem and should be further studied in future research. 

The specific contributions of this study are summarized as follows:

\begin{itemize}
\setlength{\itemsep}{0pt}
    \item We define the problem of hidden visit inference as part of the trip detection process using sparse CDR data. We show that estimated trip characteristics, such as average trip distance, would be biased without hidden visit inference.
    \item We propose a data fusion approach to obtain labeled training data from CDR data alone for supervised statistical inference of hidden visits. More specifically, labeled observations are extracted from more frequently sampled cellular data access records, and features from voice call and text messaging records.
    \item We demonstrate the proposed data fusion approach for predicting whether observed displacements contain hidden visits, and identify a range of spatial, temporal, and personal features for the prediction task. Based on a large-scale real-world dataset, we estimate that over 10\% of the displacements extracted from CDR data involve hidden visits.
\end{itemize}

The remainder of the paper is organized as follows. Based on a review of the existing literature, Section~\ref{sec:method} summarizes a process to extract trips from CDR data, and proposes a new data fusion approach for extracting labeled observations, including the specific model formulation for inferring the existence of hidden visits. A specific application with real-world CDR data is demonstrated in Section~\ref{sec:application}. Section~\ref{sec:discuss} concludes the paper with a discussion of limitations and future research.

\section{Methodology} \label{sec:method}

\subsection{Trip Extraction from CDR Data} \label{sec:lit}

Despite its increasing popularity in human mobility and transportation studies, CDR data have several limitations that hinder the ability to accurately extract individual trips. First, CDR data typically provides spatial information at the cell tower level, while the precise location of the user is unknown. It is also well documented that positioning noise exists in CDR data, which stems from signal movements \cite{calabrese_estimating_2011,iqbal_development_2014}. Low spatial resolution and signal noise both contribute to localization error. Second, the status of travel is not provided in CDR data. A mobile phone record may be generated during a visit to a place or during a trip between two places. This poses a challenge for identifying trip origins and destinations. Third, the sparse nature of CDR data makes it impossible to obtain a complete profile of user mobility. Even when complete records of the mobile phone activities of an individual are available, not every trip of the user is observable. Only those trips that occur in tandem with mobile phone activities are recoverable.

\begin{figure}[ht]
\centering
\includegraphics[width=0.9\textwidth]{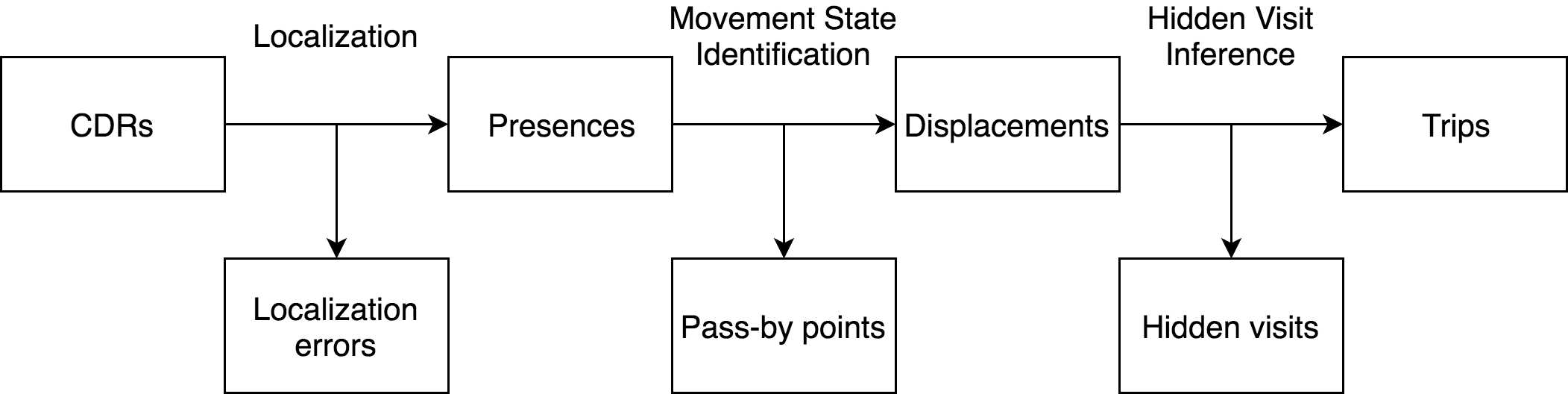}
\caption{A general process for extracting trips from CDR data}
\label{fig:trip}
\end{figure}

All the aforementioned limitations are, to various degrees, recognized and discussed in the literature. While terminologies and methodologies vary across specific studies, this section synthesizes them into a unified framework shown in Figure~\ref{fig:trip}. Generally, to extract trips from CDR data, three stages are needed---\textit{localization}, \textit{movement state identification}, and \textit{hidden visit inference}. Each is intended to address one of the three limitations. The results obtained after each stage are closer to actual individual travel behavior.

The first stage, localization, intends to mitigate localization errors and estimate user locations. A plethora of different methods have been used in the literature to reduce localization error \cite{wang_estimating_2013,isaacman_identifying_2011,csaji_exploring_2013,jiang_review_2013}. They typically include two steps---trajectory smoothing and spatial clustering. In trajectory smoothing, one takes a sequence of CDRs within a certain time threshold and applies smoothing or filtering algorithms to reduce ``jumps'' in the location sequence. These algorithms include speed-based filtering \cite{wang_estimating_2013}, time-weighted smoothing \cite{csaji_exploring_2013}, or assigning a single medoid location to every record in the sequence if they are close by \cite{jiang_review_2013}. All of them produce smoothed location sequences. In spatial clustering, one ignores the ordering or the temporal distribution of CDRs and clusters data points based on their spatial distribution only. In this way, we can consolidate points that may represent the same location but are visited on different days. Agglomerative clustering \cite{jiang_review_2013,hariharan_project_2004} and leader clustering \cite{isaacman_identifying_2011,csaji_exploring_2013} are two common spatial clustering algorithms used in prior research. The former clusters a sequence of locations based on a distance matrix only, while the latter can prioritize some locations over others usually based on the visit frequency. Cluster diameters need to be specified in both algorithms. In some cases, the location of a mobile phone may be recorded as a triangulated coordinates computed based on the locations of multiple cell towers that the device connects to. When triangulated coordinates are available, a model-based clustering method, proposed by \cite{chen_traces_2014}, is more flexible as it does not require predetermined threshold values and allows for probabilistic cluster assignments.

After localization, the location of a user at a certain time is represented by a clustered location, instead of a cell tower location. A time-stamped user location is called a \textit{presence}. Each presence can either occur during a trip or during an activity at a meaningful place. \cite{jiang_review_2013} referred to the former category as ``pass-by'' points, and the latter ``stay'' points. The goal of the second stage, movement state identification, is to distinguish between the two categories and extract visits from presences. A \textit{visit} is a series of ``stay'' points correspond to the same location. The most common way to do this in prior literature is simply to apply a dwell time threshold \cite{calabrese_estimating_2011,mellegard_origin/destination-estimation_2011,wang_estimating_2013,jiang_review_2013}. For example, a threshold of 10 min is used in \cite{jiang_review_2013}. A sequence of presences that are associated with the same location and span over 10 min is classified as a visit. Otherwise, they are flagged as pass-bys. This method works well for high-frequency data, such as GPS data. However, for sparse CDR data, it may cause most presences to be labeled as pass-by points. One way to improve this is to further identify ``potential stays'', the presences that are classified as pass-bys using the dwell time criterion but are associated with a previously visited location \cite{jiang_review_2013}.

Whereas most existing methods only cover the first two stages, we argue that a third stage, hidden visit inference, is needed to distinguish between trips and displacements. A \textit{displacement} occurs between two consecutive visits observed in the data, while a trip occurs between two consecutive visits regardless they are observed or not. In other words, a displacement may correspond to one or more trips. Using displacements extracted from sparse CDR data to directly estimate mobility patterns may lead to biases \cite{zhao_understanding_2016}. The discrepancy between displacements and trips is a non-trivial obstacle in applying CDR data for travel behavior analysis \cite{chen_promises_2016}. Hidden visit inference is a problem that has been largely overlooked in the literature. \cite{bayir_mobility_2010} is the first study that explicitly defines the problem of hidden visits. They make the distinction between ``observed end-locations'' and ``hidden end-locations''. Based on their definition, ``a hidden location occurs when a significant amount of time is elapsed during cell transition.'' In an attempt to address the issue, they propose the use of a transition time threshold to determine whether a hidden location exists during an ETI, which heavily relies on personal judgment and lacks statistical robustness. While several statistical methods have been developed to fill in the gaps in sparse CDR trajectories by estimating the length of stay at each observed location, they do not explicitly consider hidden visits to a different location \cite{hoteit_filling_2016,chen_enriching_2018}. More recently, \cite{chen_complete_2019} proposed a tensor decomposition method for complete CDR trajectory reconstruction. Specifically, a 3-dimensional tensor is constructed for each user and the missing locations are estimated based on the assumption of user behavior regularity. Unsupervised learning methods, such as tensor decomposition, are often necessary because the ground truth data about hidden visits are typically not available. However, unsupervised learning methods are generally difficult to calibrate and do not perform as well as supervised learning methods for prediction tasks. In this study, we will present a novel data fusion approach that makes it possible to infer hidden visits using supervised learning methods. In addition, unlike \cite{chen_complete_2019}, our hidden visit inference method will combine both individual-specific features and other spatiotemporal features under a universal model to allow learning across users.

\subsection{Hidden Visit Inference based on Data Fusion} \label{sec:method:fusion}

To infer whether a hidden visit exists is essentially a classification problem. It involves building a statistical model for predicting a binary output based on one or more inputs \cite{james_introduction_2013}. This requires a set of training examples, each being a pair consisting of a feature vector, $X$, and a desired output value, $Y$. However, in the case of CDR data, such training data is typically unavailable. This is arguably the most critical obstacle that limits our ability to transition from existing heuristic-based approaches to statistical approaches. 

Specifically for hidden visit inference, the complete travel profile of a user is required, along with the sparse CDR data, in order to form labeled observations. One way to achieve this is to find another data source that complements the characteristics of CDR data. CDR data are one example of large-scale urban mobility data sources that cover large user population and long observation period, but the individual-level information that is captured in such data is relatively coarse. In contrast, another type of data may be collected from a smaller sample of individuals over a shorter observation period, but can provide richer and more detailed information at the individual level, e.g., the Reality Mining dataset \cite{eagle_reality_2006}. These two types of data are complementary to each other. In this study, we refer them as \textit{coarse big data} and \textit{rich small data}, respectively. Whenever both types of data are available, we can maximize their value by combining the two for statistical learning, which involves forming training examples with $X$ extracted from coarse big data, and $Y$ from the rich small data. The trained models can then be applied to coarse big data for larger population over longer period, so that some of the unobservable information in the data can be inferred. Potential ways to collect rich small data include travel surveys and smartphone GPS tracking. Both options require active recruitment of sample users, and thus are costly and not very scalable.

Given these practical challenges, this study proposes a new way to apply data fusion using only CDR data. This is possible because multiple types of mobile phone transactions are recorded in CDR data, including voice calls, SMS, and data activities. While many of the datasets analyzed in the literature consist of voice calls only (e.g., \cite{iqbal_development_2014,gonzalez_understanding_2008,csaji_exploring_2013,barabasi_origin_2005}), or voice calls in combination with SMS records (e.g., \cite{isaacman_identifying_2011}), records of data activities are becoming more available (e.g., \cite{calabrese_urban_2014,ranjan_are_2012,calabrese_estimating_2011}). Unlike voice call and SMS activities, data activities do not always require user initiation or participation \cite{ranjan_are_2012}. On devices with enabled cellular data capability, a plethora of mobile applications, if allowed by users, make periodic or sporadic connections to the cellular network automatically. These data activities are recorded as data access records, and they tend to be less sparse than voice call or SMS records. Furthermore, voice call and SMS records are determined by mobile phone usage preferences. As a result, the mobility patterns observed from such data may be confounded with the user's mobile phone usage behavior \cite{williams_measures_2015}. On the other and, data access records can be generated passively. For example, a user may prefer not to make voice calls at certain locations or at a certain time of the day, and thus, the travel associated with these locations and periods may be hidden from the voice call records. However, these otherwise ``hidden'' visits can be captured by passively generated data access records. For these reasons, data access records can be used to capture complete travel profiles, at least for a small group of smartphone users with passively generated cellular data activities. Therefore, despite the lack of a complementary data source, it is still possible to obtain labeled data by extracting $X$ from voice calls and SMS records, and $Y$ from data access records. 

\subsection{Problem Formulation} \label{sec:method:formulation}

After localization and movement state identification, we obtain a series of stay points for each user, ${(s_1^u,t_1^u), (s_2^u,t_2^u), ..., (s_n^u,t_n^u)}$, where $s_i^u$ and $t_i^u$ are the location and timestamp of the $i$-th stay point for user $u$. A hidden visit occurs when the user makes a trip to a location other than $s_i^u$ and $s_{i+1}^u$ during the time between $t_i^u$ and $t_{i+1}^u$. It is challenging to directly estimate the location and time of the hidden visit because of the large number of possible outcomes. Instead, we focus on a simpler question in this study---whether hidden visits exist. Let $h_i^u$ indicate whether a hidden visit exists between $t_i^u$ and $t_{i+1}^u$. It is the target variable to be inferred.


For a given user, the superscript $u$ is omitted for clarity. Let $e_i$ indicate whether the time period between $t_i$ and $t_{i+1}$ counts as an ETI, i.e.,
\begin{equation}
e_i =
\begin{cases}
1, \;\; \text{if} \; t_{i+1}-t_i>\tau \\
0, \;\; \text{otherwise}
\end{cases}
\end{equation}
where $\tau$ is the minimum time threshold of ETI, which essentially defines the temporal resolution of the analysis. The choice of $\tau$ depends on the problem requirement and the data constraint. By definition, hidden visits can only exist during ETIs. Or in other words, $P(h_i=1|e_i=0)=0$. If the data is frequently sampled, $e_i=0$ $\forall i$, and no hidden visit inference is needed. Otherwise, it is necessary to estimate $P(h_i=1|e_i=1)$. One way to estimate this is to use true values of $h_i$, which may be obtained through the data fusion approach described in Section~\ref{sec:method:fusion}.

Assume that both coarse big data and rich small data are available for a group of users. A series of stay points can be obtained from the former. For each time interval $(t_i, t_{i+1})$ when $e_i=1$, a set of locations $S'_i$ are obtained from the latter. Therefore, the true values of $h_i^u$ may be determined based on the following rule:
\begin{itemize}
\item $h_i=1$, if $e_i=1$, and there exists a $s' \in S'_i$ so that $s' \neq s_i$, $s' \neq s_{i+1}$
\item $h_i=0$, otherwise.
\end{itemize}

\begin{figure}[!ht]
\centering
\includegraphics[width=0.6\textwidth]{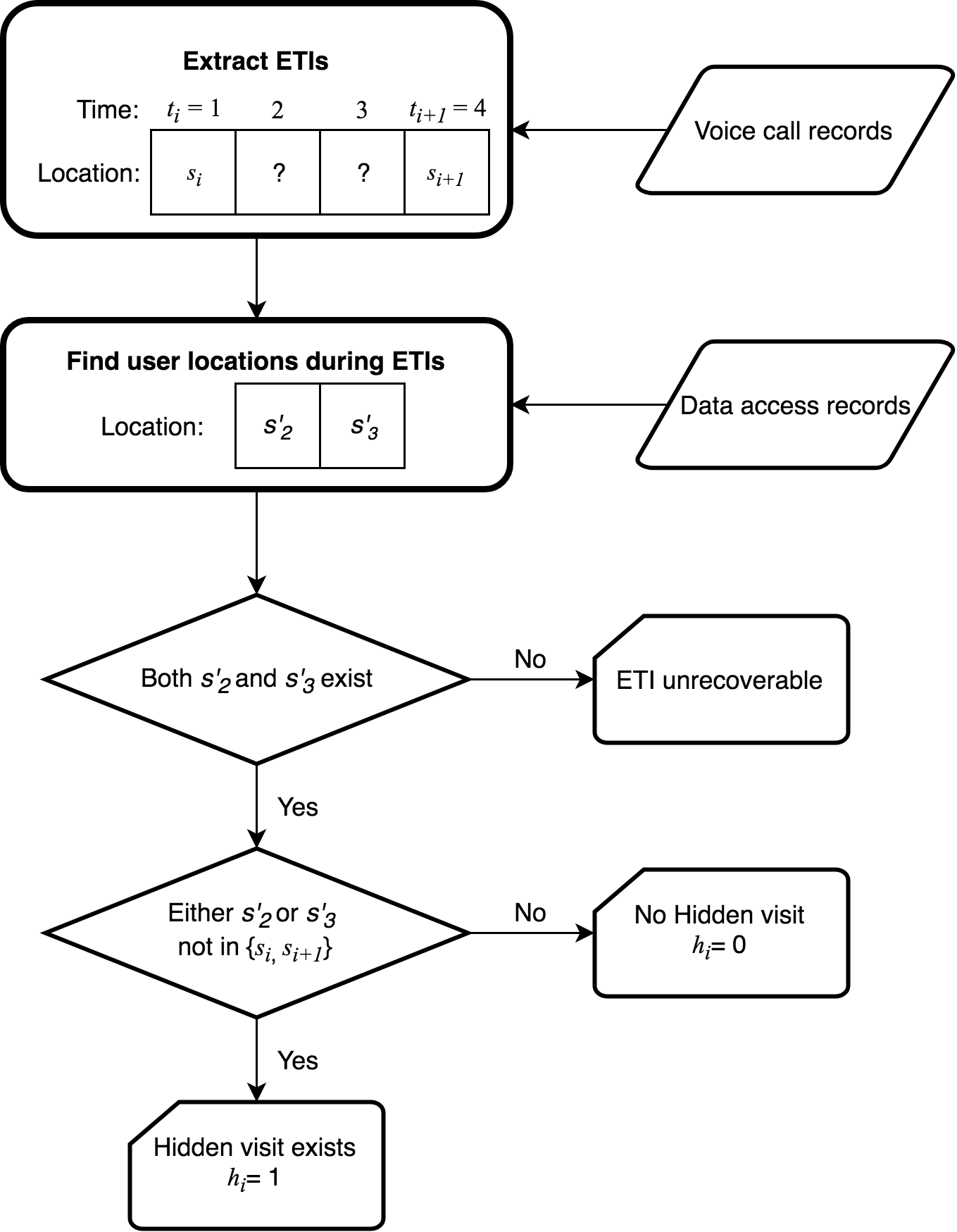}
\caption{Illustration of the process for identifying the existence of hidden visits}
\label{fig:fusion}
\end{figure}

To ensure that the location sequences in the two data sources are comparable, we may transform them into discrete time series, for example, by binning the timestamps into hours. Also, in reality, even in the frequently sampled data access records, user locations may be missing in certain periods. For example, if the mobile phone of a user is out of battery for a period, all the travel activities during the period would be missing. We call these periods unrecoverable. Only hidden visits within the recoverable ETIs can be identified. This process of obtaining labeled observations is illustrated using the example in Figure~\ref{fig:fusion}.

With labeled observations, a model can be trained to estimate $P(h_i=1|e_i=1)$. For parametric methods, the training data are used to find the values of a set of model parameters, $\theta$, so that some loss function is minimized. Assuming there are $M$ users in the training data, each with $N_u$ observations, the objective function should be
\begin{equation}
\hat{\theta} = \text{argmin} \sum_{i=1}^N e_i L(h_i,f(X_i; \theta))
\end{equation}

where $N$ is the number of observations in the training data, $X_i$ is the feature vector of $i$-th observation, and $L(a,b)$ is the loss function specified by the true value, $a$, and the estimated value, $b$. The specific form of the loss function and $f(X_i^u; \theta)$ depends on the choice of algorithms.

It is important to consider two different scenarios depending on whether $s_i = s_{i+1}$. In Scenario I, $s_i \neq s_{i+1}$. A displacement occurs from $s_i=A$ to $s_{i+1}=B$, and the goal of hidden visit inference is to determine whether the user visits another place $Z$ in between, e.g., $A \rightarrow Z \rightarrow B$. In Scenario II, $s_i = s_{i+1}$. In this case the goal is to determine whether the user visits another place and returns, e.g., $A \rightarrow Z \rightarrow A$. The two scenarios have different implications regarding user behavior and travel patterns, and thus require separate model specifications, even though the general methodology may be similar. This paper focuses on Scenario I and demonstrate how the data fusion approach may be implemented to infer $P(h_i=1|e_i=1,s_i \neq s_{i+1})$.

\section{Application} \label{sec:application}

\subsection{Data} \label{sec:data}

The dataset used for this study is collected by one of the major cellular service operators from a Chinese city with a population of 6 million. The dataset contains over 2 billion mobile phone transaction records generated by 3 million users during November 2013. Only voice call and data access records are available; SMS records are not available, exacerbating the sparsity problem at the individual level. The key fields in the CDR data include:
\begin{itemize}
\item ID - encrypted unique identifier for each phone number
\item Location Area Code (LAC) - location area code, used in combination with Cell ID to identify the cell tower used for the transaction
\item Cell ID - used in combination with LAC to identify the cell tower used for the transaction
\item Date Time - the timestamp of the mobile phone transaction
\item Event ID - the type of the event that triggers the transaction, which may be an outgoing call, incoming call, or data usage (2G/3G).
\end{itemize}

In addition to the CDR dataset, we have a cell tower database documenting the attributes (including geographic location) of the cell towers. This makes it possible to query the coordinates (in the form of longitude and latitude) of the tower associated with each mobile phone record using LAC and Cell ID. In total, we are able identify the locations of over 9,000 cell towers that appear in the CDR dataset. 

\begin{figure}[ht]
\centering
\includegraphics[width=0.8\textwidth]{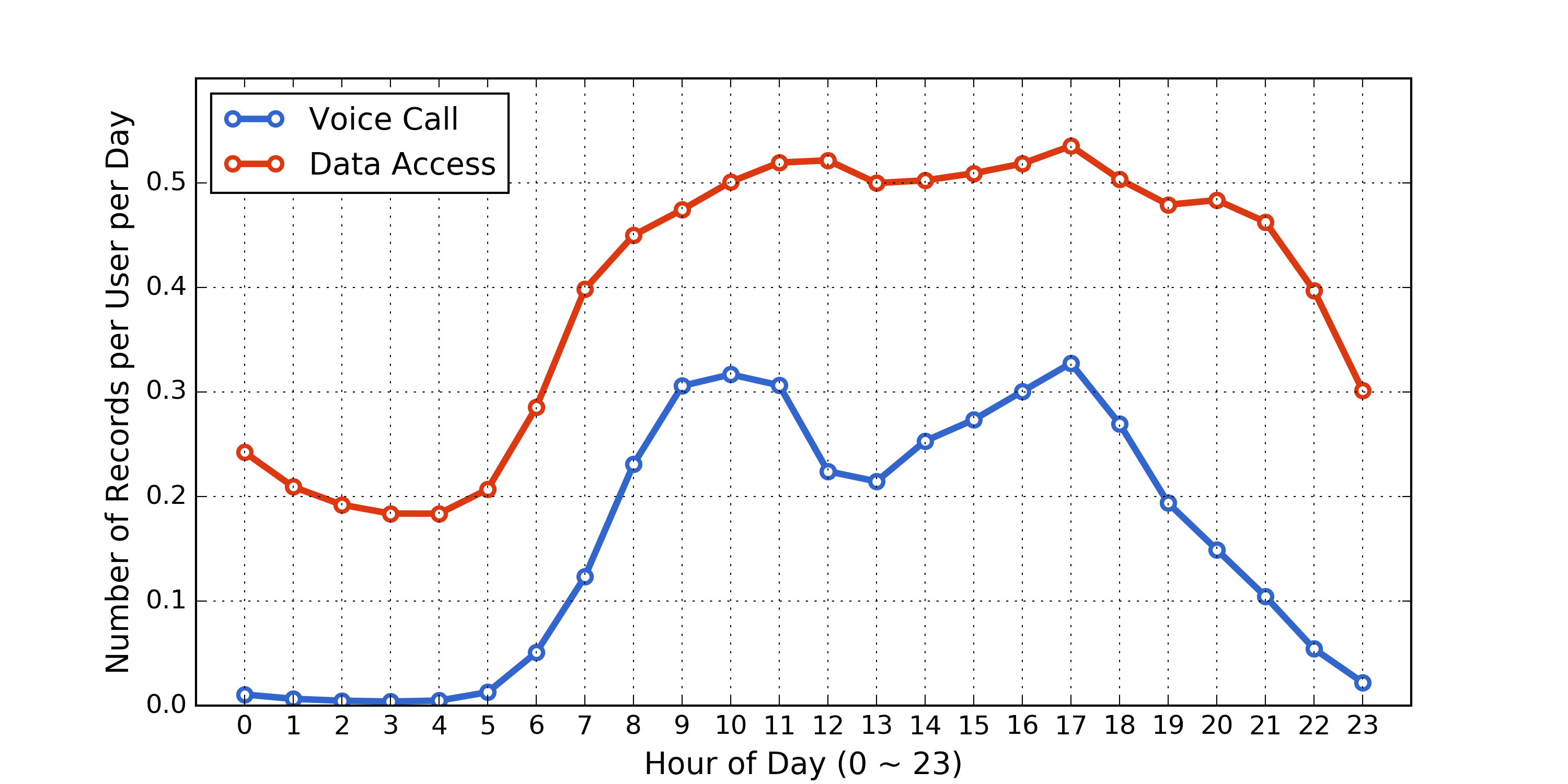}
\caption{Mobile phone usage pattern over time of day}
\label{fig:cdr}
\end{figure}

Although the total amount of CDR data is large, the number of records per user is sparse. On average, a user generates 0.16 voice call records and 0.40 data access records every hour. Voice call and data access records exhibit different patterns. As shown in Figure~\ref{fig:cdr}, data access records are not only larger in number but also more evenly distributed throughout the day than voice call records. Similar to prior findings in \cite{candia_uncovering_2008}, the number of voice call records per user have two peaks, one in the morning and the other in the afternoon, which resembles the distribution of travel demand. It suggests that making phone calls is somewhat correlated with travel, potentially causing biases in travel estimation. For example, if some users only make phone calls before and after their commutes, we may overestimate the proportion of commuting trips and underestimate other trips. Without observing the actual travel behavior from another less biased data source, it is very difficult to correct the bias. Data access records suffer from a similar problem, but to a lesser degree. Therefore, data access records may be used to quantify, and potentially mitigate, the biases of voice call records.

\begin{figure}[ht]
\centering
\includegraphics[width=0.8\textwidth]{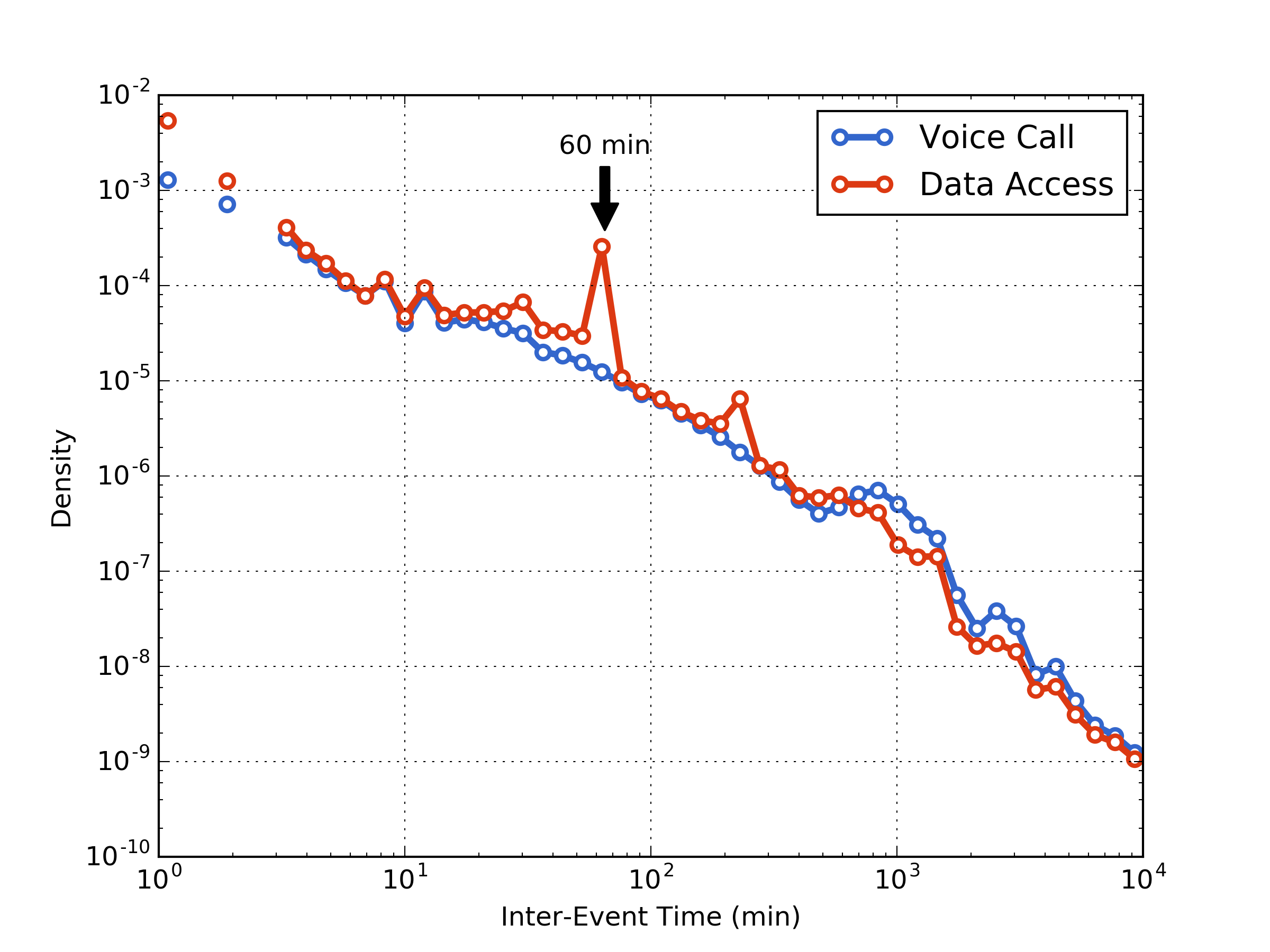}
\caption{Distribution of inter-event time by transaction type}
\label{fig:interevent}
\end{figure}

The inter-event distribution of the voice call and data activities is also explored. The inter-event time is calculated as the time difference between two consecutive records for the same user. Again, voice call records and data access records are analyzed separately. Figure~\ref{fig:interevent} shows that, whereas the inter-call time is characterized by a smooth distribution curve, the inter-event time for data activities exhibits a few peculiar spikes, the most significant of which being at the 1-hour mark. This is likely caused by the fact that some mobile applications are set up to automatically make hourly connections to the cellular network. This finding suggests that a reasonable choice for ETI threshold $\tau$ is 1 hour, at least for this dataset.

Note that not every user's data access records have such regular hourly inter-event intervals. Some users may not have a smartphone, while others may disallow for passive data usage of some mobile applications. An examination of mobile phone usage patterns shows that 9\% of the users have only data access records but no voice calls, possibly because they represent tablets or secondary devices. The rest of the users are considered regular mobile users who generated at least one voice call record in November. They can be further divided into call-only users who do not use cellular data (43\%), and mixed users who have both voice call and data access records (57\%).

\begin{figure}[ht]
\centering
\includegraphics[width=0.8\textwidth]{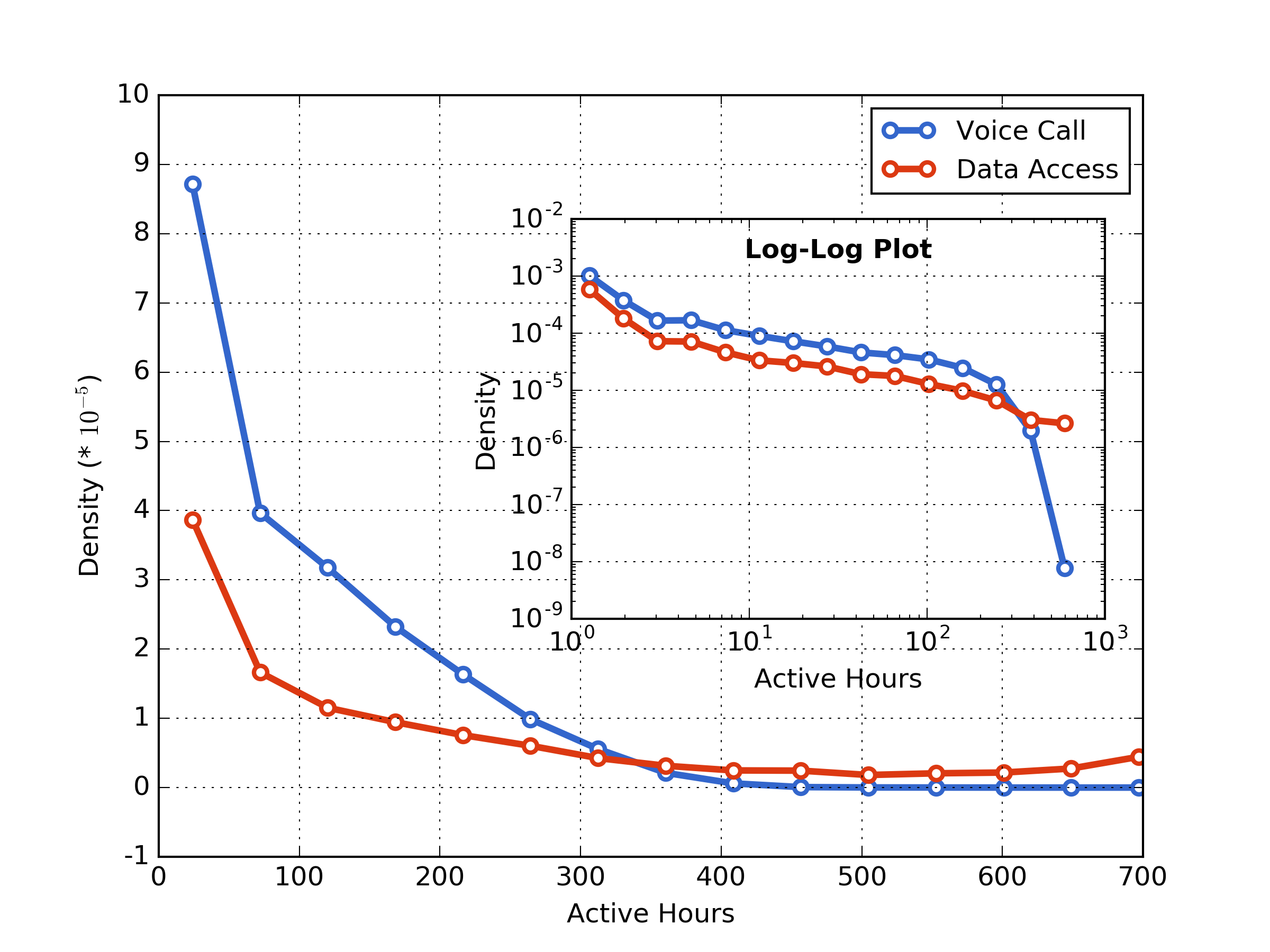}
\caption{Distribution of mixed users by number of active hours}
\label{fig:active}
\end{figure}

By further decomposing mixed users, we find that a large proportion of data access records are generated by a small group of users. Figure~\ref{fig:active} shows the distribution of the active hours of mixed users based on their voice call or data access frequency. An active hour denotes an hour when the user generates at least one record, and we break this down by the two transaction types. Note that there are 720 hours in total during our study period (i.e., 30 days). The two distributions shown in the figure are distinctly different, which is highlighted in the log-log plot (see the inset chart of Figure~\ref{fig:active}). There is virtually no user that has more than half of their hours with at least one voice call, but there are a small group of users that generate data access records in most of the hours, a strong indication that these users have passively generated data activities. In this study, we define \textit{frequent data users} as the mixed users who have active data usage in at least half of the hours (in this case, 360 hours). The threshold represents a trade-off between certainty and volume---a lower threshold would place more users in this group, but we would be less certain that these users have passively generated data access records. For this group of users, their voice call records are still sparse, but their data access records are not. Therefore, we can to some extent observe their mobility between voice calls based on their data usage. Note that the frequent data users only account for 10\% of the regular mobile user population. Nevertheless, hidden visit inference models may be trained based on CDRs of the frequent data users, before being deployed for the large majority of users with sparse CDR data. Note that even a frequent data user may still have ETIs in their trajectories, e.g., when the user is out of cellular coverage or is served by unknown Wi-Fi hotspots. We do not require them to have complete trajectories. Instead, we only extract the complete segments of trajectories for training data. After the model is trained, it can then be deployed for hidden visit inference for all ETIs.

\subsection{Preprocessing}

Before hidden visit inference, two previous stages need to be completed---localization and movement state inference. For localization, a method similar to \cite{isaacman_identifying_2011,csaji_exploring_2013} is adopted to perform spatial clustering of cell towers and identify important user locations. The method has two steps. In the first step, the cell towers are ranked based on their importance for each user, where the importance is measured based on the number of ``call-days'' (i.e., days the cell tower was contacted). In the second step, the leader algorithm is used for clustering analysis. The algorithm starts with the most important cell towers and merges the surrounding towers into the first cluster. Then we move on to the most frequent tower of the remaining towers and repeat, until all towers are assigned a cluster. Unlike the K-means algorithm, the leader algorithm does not require a predefined number of clusters, which is advantageous because localization is performed at the individual level. Compared to the hierarchical clustering algorithm, the leader algorithm allows us to assign higher priority (or weight) to the frequently used cell towers. The weighted centroid of the cell towers with a cluster is used to represent the location of an user.

In the movement state identification stage, we distinguish between stay and pass-by points based on the dwell time and the frequency of appearances at the associated user location. To determine dwell time, presences are aggregated to segments based on location matching and temporal proximity. Two consecutive presences are combined if they are associated with the same location ($s_i=s_{i+1}$) and the time difference is within the ETI threshold ($t_{i+1}-t_i<\tau$). Each presence segment is associated with a single location and covers a unique time span. If a segment's time span is above the minimum dwell time threshold (set as 10 min \cite{jiang_review_2013}) or its associated location appears at a frequency higher than the minimum frequency threshold (set as 4 days per month), all the presences in the segment are classified as stay points; otherwise, they are flagged as pass-by points and will not be used for further analysis. Here, we assume that a location is an important place for users if they appear at the location at least once a week. This is loosely based on \cite{isaacman_identifying_2011}, which defines an important place as ``a geographic location where a person spends a significant amount of time and/or which she visits frequently''.

For model implementation, 10,000 frequent data users are selected to form training samples. For each user in the dataset, we are able to obtain a series of stay points $(s_1,t_1)$, $(s_2,t_2)$, ..., $(s_n,t_n)$ extracted from voice call records, and corresponding hidden visit labels $h_1, h_2, ..., h_{n-1}$ extracted from the data access records based on the approach discussed in Section~\ref{sec:method:formulation}. For simplicity, we filter the data to only keep the observations where $e_i=1$ and $s_i \neq s_{i+1}$. In other words, each observation is a displacement with ETI. A displacement with ETI is characterized by two consecutive stay points $(s_i,t_i)$ and $(s_{i+1},t_{i+1})$, where $s_i \neq s_{i+1}$ (thus a displacement) and $t_{i+1}-t_i>\tau$ (thus an ETI). $h_i$ indicates whether there exists a hidden visit during the ETI. For example, if two consecutive stay points are (A, 8 am) and (B, 11 am), it counts as a displacement with ETI, and the goal is to infer whether there exits a hidden visit between 8 am and 11 am. For model training, we only consider instances when $h_i$ can be recovered from the data access records. For each user, we randomly select one observation. This gives us a dataset of 9,761 observations, which is then used for model training. As this is a binary classification problem, the performance of classifiers are evaluated using standard metrics such as classification accuracy, precision, recall, F1 score, and the area under the receiver operating characteristic curve (ROC AUC). The precision and recall values are calculated based on the ``positive'' class, which is the class with $h_i=1$. 10-fold cross-validation is used to obtain robust model performance metrics.

\subsection{Model Specification} \label{sec:model}

Given a displacement with ETI, a set of attributes need to be defined in order to be used for hidden visit inference. Generally, its attributes can be categorized into three sets---spatial (of the displacement), temporal (of the ETI) and personal (of the user) attributes.

Spatial attributes refer to the characteristics of the displacement $(s_i,s_{i+1})$. Distance is a measure of travel cost between $s_i$ and $s_{i+1}$, and the proportion of observed trips (i.e., displacements with no ETIs) is a measure of the strength of connection. In an attempt to characterize locations, the home and workplace of an individual are inferred based on simple heuristics. For each user, we identify the top two most frequently visited locations (a minimum frequency threshold has to be met), and determine that the one with more presences during (1) all hours on weekends and (2) night hours (i.e., before 7 am or after 7 pm) on weekdays is home \cite{alexander_origindestination_2015} and the other the workplace. The term ``workplace'' is used to generally indicate the location that an user visits frequently during daytime on weekdays. It may be a school for some users. This categorization allows us to semantically characterize displacements in a few ways. Based on the function of only $s_{i+1}$, we may categorize displacements by three travel purposes---home, work, or other. Alternatively, displacements may be grouped into four categories based on the functional combination of both $s_i$ and $s_{i+1}$---home-based work (HBW), home-based other (HBO), work-based other (WBO), and other-based other (OBO). HBW is travel between home and work, HBO between home and other, WBO between work and other, and OBO between two other locations. This categorization is commonly used in transportation planning. In addition, we also examine the distribution of the user's displacements without ETIs and see how many of them are between $s_i$ and $s_{i+1}$. 

Temporal attributes refer to the characteristics of the ETI $(t_i,t_{i+1})$. The duration of ETI, or $(t_{i+1}- t_i)$, is a important factor. In general, the longer the ETI, the more likely there exists a hidden visit. To determine the time of day effect, we use four dummy variables: whether the ETI overlaps with (i) morning peak hours (from 7 am to 9 am), (ii) afternoon peak hours (from 4 pm to 7 pm), (iii) midday hours (from 10 am to 3 pm), and (iv) night hours (from 8 pm to 6 am the next day). These dummy variables are not mutually exclusive because ETIs often span across multiple hours. The underlying assumption is that people have different motility patterns during different periods in a day. In addition, we calculate how often a user appears at other locations (neither $s_{i}$ nor $s_{i+1}$) during the same time of the day (TOD) as the ETI $(t_i,t_{i+1})$. For example, if the ETI is between 8am and 11am, we will count the total number of hours where the user is observed elsewhere between 8am and 11am across all days in the observation period.

User attributes include both characteristics of mobile phone usage and those related to travel behavior. One specific measure of travel tendency is the number of displacements per active hour, which is a normalized measure of user displacement rate. Table~\ref{tab:feature} presents numerous attributes that are extracted and tested, and the italicized ones are those that are selected to be included in the final model based on model validation.

\begin{table}[ht]
  \centering \small
  \caption{Possible features for hidden visit inference}
  \begin{threeparttable}
    \begin{tabular}{c p{10cm}}
    \toprule
    Category & Features \\
    \midrule
    \multirow{6}{*}{Spatial Attributes} & \textit{Distance between $s_{i}$ and $s_{i+1}$} \\
    & \textit{Displacement type}* \\
    & Location ranking of $s_{i}$ and $s_{i+1}$* \\
    & Visit frequency of $s_{i}$ and $s_{i+1}$* \\
    & \textit{Location function of $s_{i}$ and $s_{i+1}$ (home, work, others)}* \\
    & \textit{\% of displacements (without ETIs) between $s_{i}$ and $s_{i+1}$}* \\
    \midrule
    \multirow{6}{*}{Temporal Attributes} & \textit{Duration} \\
    & \textit{Time of the day (TOD)} \\
    & Day of the week \\
    & Number of locations where user appears during same TOD* \\
    & \textit{Frequency of user appearing elsewhere (neither $s_{i}$ nor $s_{i+1}$) during same TOD}* \\
    \midrule
    \multirow{6}{*}{User Attributes} & Number of voice calls* \\
    & Number of active call hours* \\
    & \textit{Number of visited locations*} \\
    & Number of visited locations per active call hour* \\
    & Number of displacements* \\
    & \textit{Number of displacements per active call hour*} \\
    \bottomrule
    \end{tabular}%
    \begin{tablenotes}
      \small
      \item Note: features with * are derived using individual user data; features in \textit{italics} are included in final models.
    \end{tablenotes}
  \end{threeparttable}
\label{tab:feature}%
\end{table}%

Because different assumptions for loss functions and model structures may yield different results, four commonly used classifiers are tested. They are logistic regression, support vector machine (SVM), random forest, and gradient boosting. The implementation details of these methods are described as follows:
\begin{itemize}
\item The logistic regression model outputs the probability distribution across the two classes, and thus a cut-off value needs to be chosen to produce an point estimate (i.e., yes or no). Based on preliminary tests, the cutoff value is set at 0.5. To avoid overfitting, the L2 regularization is used, and the parameter of inverse regularization strength, $C$, is chosen to be 1.0.
\item For SVM, the radial basis function kernel is used. It takes the following form: $K(x,x')=\text{exp}(-\gamma||x-x'||^2)$, where the parameter, $\gamma$, needs to be specified based on the validation set. Based on preliminary tests, it is determined that $\gamma=0.05$. Similar to logistic regression, the regularization parameter $C=1.0$.
\item A random forest is an ensemble method that fits a number of decision tree classifiers on various sub-samples of the dataset and use averaging to improve the predictive accuracy and control over-fitting. The Gini index is used to measure the impurity of a node in a tree, and the number of trees in the forest is set to be 50. 
\item Gradient boosting is an ensemble method that builds an additive model in a forward stage-wise fashion and allows for optimization of an arbitrary differentiable loss function \cite{friedman_greedy_2001}. In each stage, a weak model, typically a decision tree classifier, is fitted based on the negative gradient of the loss function.
\item All classifiers are implemented in Python through the machine learning package \textit{scikit-learn} \cite{pedregosa_scikit-learn:_2011}. If not specified, the default model settings are used.
\end{itemize}

\subsection{Feature Importance Analysis}

Of the classifiers used, logistic regression has most interpretable model parameters. Thus, the detailed results of the logistic regression model are presented in Table~\ref{tab:logistic} for a better understanding of the relationship between variables. Positive coefficients mean that an increase in attribute values will increase the probability that a hidden visit occurs, and vice versa.

\begin{table}[ht]
  \centering \small
  \caption{Logistic regression model results}
    \begin{tabular}{p{4in}rr}
    \toprule
    Feature & Estimate & p-value \\ 
    \midrule
    Intercept & -1.683 & 0.000 \\
    \textbf{Spatial Features} & & \\
    Distance between $s_i$ and $s_{i+1}$ (in km) & 0.087 & 0.000 \\
    Destination = Home & 0.169 & 0.011 \\
    Displacement type = HBW & -0.980 & 0.000 \\
    Displacement type = HBO & -0.552 & 0.000 \\
    Displacement type = WBO & -0.415 & 0.000 \\
    If any trip from $s_i$ to $s_{i+1}$ is observed & -0.270 & 0.000 \\
    \% of displacements (without ETIs) between $s_{i}$ and $s_{i+1}$ & -0.627 & 0.037 \\
    \textbf{Temporal Features} & & \\
    Duration of the ETI (in hours) & 0.031 & 0.000 \\
    ETI overlaps with morning peak hours & 0.190 & 0.016 \\
    ETI overlaps with afternoon peak hours & 0.136 & 0.027 \\
    ETI overlaps with midday hours & 0.119 & 0.040 \\
    If user appears in other locations during same TOD & 0.249 & 0.003 \\
    Frequency of user appearing elsewhere during same TOD & 0.034 & 0.000 \\
    \textbf{Personal Features} & & \\
    Number of observed user locations & -0.009 & 0.015 \\
    Number of displacements per active hour & 1.177 & 0.000 \\
    \midrule
    \textbf{Model Fit} & & \\
    \multicolumn{3}{c}{chi-square = 2078.75, degrees of freedom = 16, McFadden's $R^2$=0.162} \\
    \bottomrule
    \end{tabular}%
  \label{tab:logistic}%
\end{table}%

As expected, spatial attributes matter in the classification problem. Hidden visits are more likely to occur when the displacement distance is longer. Interestingly, they also occur more often when the displacement ends at home. One possible explanation is that people are more likely to make a short visit to another place (e.g., grocery store) on their way home. This may be because that people have fewer time constraints when they travel back home. If either $s_i$ or $s_{i+1}$ correspond to the home or workplace, the probability of a hidden visit decreases. This may be caused by the fact the home and the workplace are the two most frequently visited locations for a user, and the probability of visiting any other location is relatively small. If the user is observed to travel from $s_i$ to $s_{i+1}$ on other occasions (in cases of displacements with no ETIs), there is less likely to be a hidden visit. Human mobility has been found to show high regularity \cite{song_limits_2010}. People tend to repeat the same trip over time.

In terms of temporal attributes, the ETI duration has a similar effect as the displacement distance; the longer the ETI, the more likely a hidden visit occurs. If the ETI overlaps with morning peak hours, it is more likely to involve hidden visits. Afternoon peak and midday hours have similar, but lesser, effects, while the coefficient for night hours is insignificant. This finding matches our expectation that people are generally more mobile in peak hours than at midday, and least active during night hours. If the user is observed to appear in other locations during the same period on other days, hidden visits are also more likely to occur.

The number of displacements per active hour approximates the travel rate of a user. As expected, a user with a higher travel rate is more likely to undertake a hidden visit. We find that, although the number of user locations per active hour is not a significant factor in the model, the total number of observed user locations is. This suggests that the more visited locations revealed in the data, the less likely the user has hidden visits. One may argue this is a result of call frequency; a frequent caller reveals more visited locations. However, the frequency of the call activities is also insignificant in the model. A more likely explanation is rooted in the user's spatial preference regarding phone calls. Regardless of call frequency, some users distribute their phone calls across all locations, while others may prefer to make phone call only at a few locations. The latter group of users is more likely to have hidden visits in their voice call records. In other words, the distribution of calls matters more rather than the frequency of calls.

Another way to assess feature importance is to see how much they contribute to the actual prediction of hidden visits. To do this, we evaluate the overall prediction performance, measured by classification accuracy and ROC AUC. Then, we remove spatial, temporal, and personal features from the model, and compare the resulting difference in prediction performance. The results are summarized in Table~\ref{tab:prediction}. Based on the results, it seems that spatial features are most important (because of the largest drop in prediction performance), followed by temporal features. Personal features are least important.

\begin{table}[ht]
  \centering \small
  \caption{Comparison of Prediction Performance with Different Features}
    \begin{tabular}{lcc}
    \toprule
    Feature Combination & Accuracy & ROC AUC \\
    \midrule
    Spatial + Temporal + Personal & 0.694 & 0.696 \\
    Spatial + Temporal & 0.694 & 0.695 \\
    Spatial + Temporal & 0.668 & 0.675 \\
    Temporal + Personal & 0.621 & 0.626 \\
    \bottomrule
    \end{tabular}%
  \label{tab:prediction}%
\end{table}%

\subsection{Comparison of Model Performance}

Table~\ref{tab:result} shows the performance of the four classifiers. They are compared against two baseline models. Baseline 1 is a deterministic model that assumes no hidden visit, which is the assumption that many prior studies have made when they extracted trips from CDR data. It reaches a classification accuracy of 62.8\%, meaning that the naive rule will underestimate the number of trips by at least 37\%. Note that the precision and recall are both 0 for Baseline 1, because there are no positive cases predicted. Baseline 2 is a probabilistic model that generates predictions through sampling the marginal distribution of the target variable observed in the training data. Its prediction accuracy is lower, but, unlike Baseline 1, it produces positive precision and recall. Compared to the baseline models, the fitted statistical models significantly improve the prediction performance on all metrics. Among the four classifiers, gradient boosting performs best in terms of overall accuracy and precision, while logistic regression does better in recall, F1 score, and ROC AUC. Because the positive class accounts for a smaller proportion than the negative class, it tends to be under-classified, resulting in a lower recall than precision. Common strategies to address the data imbalance include oversampling or overweighting the positive class, or, for probabilistic models such as logistic regression, adjust the cut-off probability threshold. However, these strategies may worsen the overall accuracy score. This is a trade-off to be assessed depending on specific applications.

\begin{table}[ht]
  \centering \small
  \caption{Comparison of Classification Performance}
    \begin{tabular}{lccccc}
    \toprule
    Methods & Accuracy & Precision & Recall & F1 Score & ROC AUC \\
    \midrule
    Baseline 1 (deterministic) & 0.628 & 0.0 & 0.0 & N/A & N/A \\
    Baseline 2 (probabilistic) & 0.537 & 0.366 & 0.363 & 0.365 & 0.499 \\
    Logistic Regression & 0.694 & 0.566 & 0.702 & 0.627 & 0.696 \\
    SVM & 0.712 & 0.652 & 0.459 & 0.538 & 0.659 \\
    Random Forest & 0.721 & 0.654 & 0.503 & 0.569 & 0.675 \\
    Gradient Boosting & 0.729 & 0.671 & 0.511 & 0.580 & 0.683 \\
    \bottomrule
    \end{tabular}%
  \label{tab:result}%
\end{table}%

\begin{figure}[ht]
\centering
\includegraphics[width=0.8\textwidth]{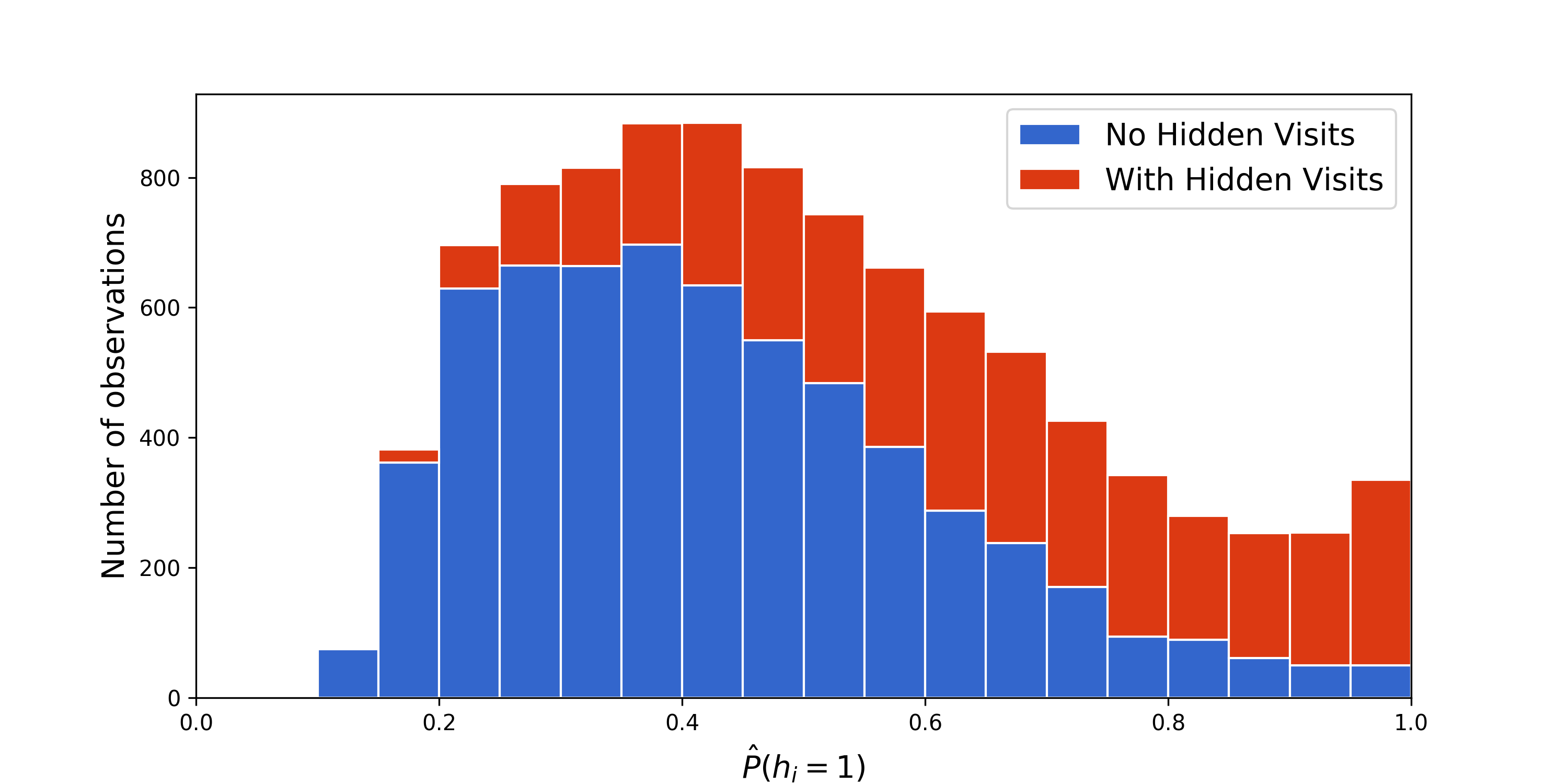}
\caption{Proportions of positive/negative classes varying by estimated probabilities of hidden visits}
\label{fig:lr}
\end{figure}

Many classifiers, such as logistic regression, can produce probabilistic predictions for a new observation. Standard SVM does not provide such probabilities, but it can with Platt scaling \cite{platt_probabilistic_1999}. Figure~\ref{fig:lr} shows how the proportions of positive (in red) and negative (in blue) classes vary based on $\hat{P}(h_i=1)$ estimated by the logistic regression model. Generally, as $\hat{P}(h_i=1)$ increases, the relative proportion of the positive class rises. However, when $\hat{P}(h_i=1)$ is between 0.5 and 0.7, the model can not confidently distinguish between the two classes. Instead, we may directly use $\hat{P}(h_i=1)$ as an indicator of uncertainty. In many applications, especially at the aggregate level, probabilistic predictions are preferred over point predictions, because they directly account for the degree of uncertainty in the analysis. This is a distinct advantage of statistical models over heuristic-based approaches, which cannot provide reliable probabilistic estimations.

\subsection{Model Deployment for Trip Extraction}

To demonstrate the importance of the hidden visit inference model, we apply the logistic regression model to all displacements with ETIs (with unknown $h_i$) in the CDR dataset. To avoid the high uncertainty associated with very long periods without observations, we only focus on the displacements with durations shorter than 8 hours. 56.5 million such displacements are extracted from the dataset, of which 23.2 million (or 41\%) are associated with ETIs. Based on the calculated probabilities $\hat{P}(h_i=1)$ from the model, we estimate that the expected number of displacements with hidden visits is over 6.4 million---27.7\% of the displacements with ETIs, or 11.4\% of all displacements. 

\begin{figure}[ht]
\centering
\includegraphics[width=0.8\textwidth]{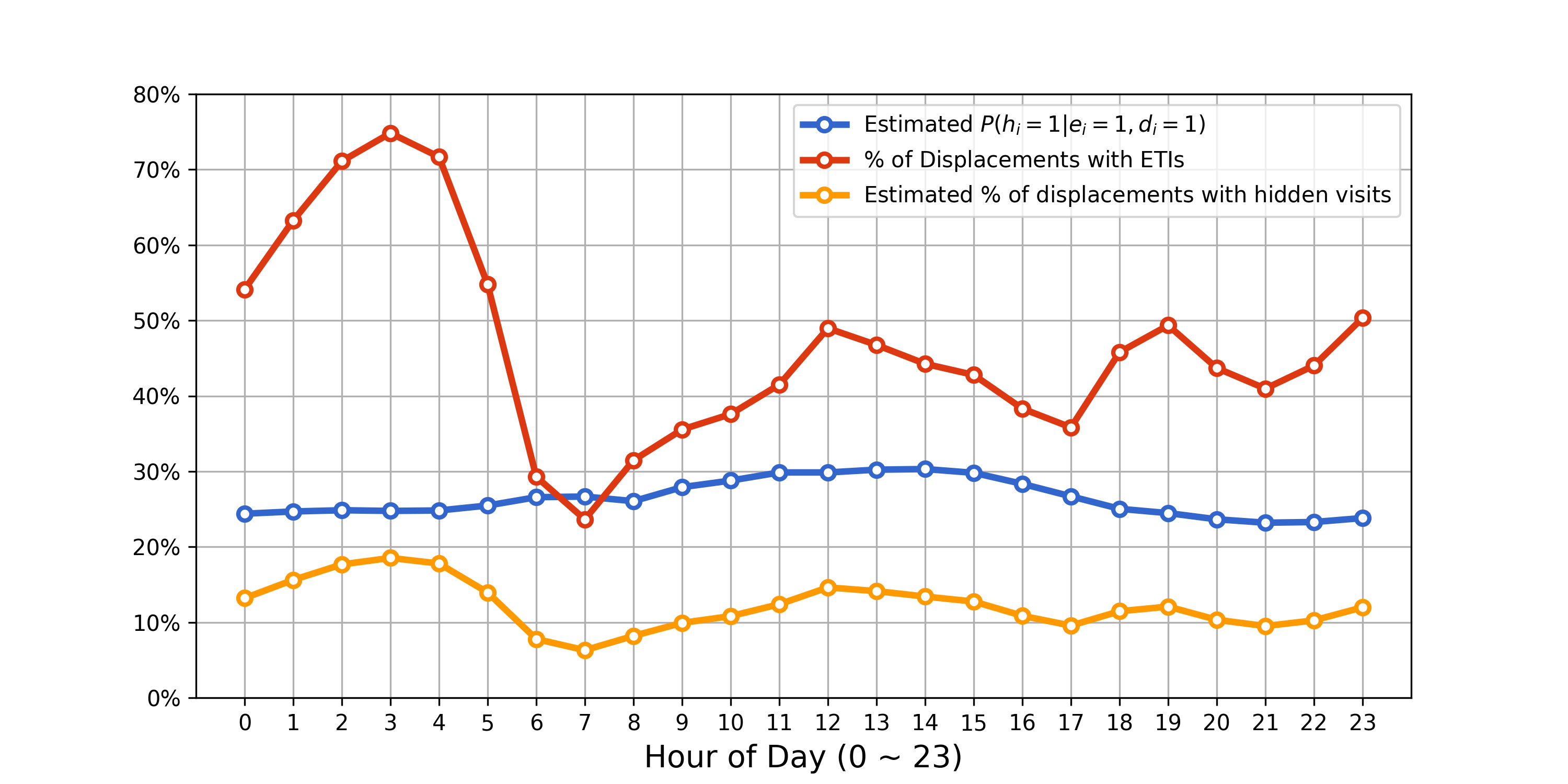}
\caption{Percentage of inferred hidden visits over time of day}
\label{fig:hidden}
\end{figure}

This has two implications. On the one hand, the results show that more than 10\% of the displacements are not direct trips, and considering their end locations $(s_i,s_{i+1})$ as OD pairs would be inaccurate. Thus, in order to obtain an accurate estimation of OD distribution, we may discount those displacements with hidden visits. On the other hand, each displacement with a hidden visit corresponds to more than one trip, and as a result the actual number of trips is at least 11.4\% more than the number of observed displacements. One way to correct this underestimation is to apply an upscaling factor based on the estimated proportion of hidden visits. Note that this proportion varies significantly over time, as shown in Figure~\ref{fig:hidden} as the orange curve. Higher proportions occur during early mornings and middays. The variation is determined by two components---the variation in the proportion of displacements with ETIs over all displacements, and the estimated $P(h_i=1|e_i=1,s_i \neq s_{i+1})$, both of which are also shown in Figure~\ref{fig:hidden} as the blue and red curves, respectively. The former is a result of the uneven distribution of CDRs over time of day (see Figure~\ref{fig:cdr}), and the latter is estimated by the hidden visit inference model. Figure~\ref{fig:hidden} suggests the peak in the early morning is primarily driven by the former component, while both components contribute to the higher percentage of hidden visits around middays. It is intuitive that during the periods with less mobile phone activity, there are greater chances of encountering ETIs, and thus, hidden visits are more likely to occur. Note that this is not to say there are more hidden visits occurring in the early morning. Instead, the correct way to interpret this is that each displacement in the early morning is more likely to contain a hidden visit. These variations need to be taken into consideration when we upscale the number of displacements to number of trips.

As another demonstration of the value of hidden visit inference model, we compare the average trip distance for displacements with or without ETIs. Average trip distance is an important indicator of travel demand, useful for transportation planning. Hypothetically, the presence of ETIs should not significantly alter the average trip distance. However, because of potential hidden visits during ETIs, displacements with ETIs are more likely to contain more than one trips, resulting in longer average distance than direct trips. As shown in Table~\ref{tab:distance}, the displacements with ETIs have much longer distance than displacements without ETIs, likely as a result of hidden visits. To address the inconsistency, we adopt the logistic regression model for hidden visit inference, and use the predicted probability of no hidden visit $\hat{P}(h_i=0|e_i=1,d_i=1)$ as the weight for each displacement. As a result, the weighted average distance is more consistent with displacements without ETIs. Note that longer displacements with ETIs are more likely to involve hidden visits, and thus they are more likely to be down-weighted. Therefore, the average distance is lower after weighting.

\begin{table}[ht]
  \centering \small
  \caption{Comparison of Average Distance across Different Types of Displacements}
    \begin{tabular}{lc}
    \toprule
    Type of Displacements & Average Distance (km) \\
    \midrule
    Displacements without ETIs & 4.489 \\
    Displacements with ETIs & 7.551 \\
    Displacements with ETIs weighted by $\hat{P}(h_i=0|e_i=1,d_i=1)$ & 4.879 \\
    \bottomrule
    \end{tabular}%
  \label{tab:distance}%
\end{table}%

\section{Discussion} \label{sec:discuss}

In this study, we define the problem of hidden visits caused by data sparsity, and develop a data fusion approach to infer the existence of hidden visit in CDR data. The proposed method works by extracting labeled observation from more granular cellular data access records and features from voice call and/or SMS records. It is demonstrated using the CDR data of 3 million users from a large Chinese city over a one-month period. The records of a sample of 10,000 frequent data users are used to train hidden visit inference models. The test results show that the developed models offer superior performance compared to the implicit assumption of no hidden visit adopted by many prior studies. Furthermore, it allows us to explicitly account for uncertainty in hidden visit inferences via probabilistic estimates. In addition, the results reveal that longer displacements are more likely to involve hidden visits. By applying the trained model to general user population, we find 11.4\% of the 56.5 million displacements extracted from CDR data involve hidden visits. This means, without considering hidden visits, the trip distance estimated from CDR data may be over-estimated and more than 10\% of the observed OD pairs are potentially inaccurate. These findings provide a better understanding regarding the potential biases of sparse CDR data, especially voice call records, for travel estimation.

The proposed methodology presents a promising research direction, and opens up many opportunities for future advancement. First, the presence of signal noises, or localization errors in particular, in CDR data limits the performance of hidden visit inference models. Better localization methods can further reduce signal noises and potentially improve the performance of the models. Second, incorporating more features and sequential dependence can help improve the performance. Mining of individual-level longitudinal data may reveal more features regarding the user's activity patterns and routines. Sequential dependence exists across a series of displacements of the same user, and it may be accounted for using methods like conditional random field. Third, as the models are developed based on training samples extracted from frequent data users' CDRs, it is assumed that the model parameters are applicable to the general user population. The validity of the assumption depends on the problem and model specifications. Future research is needed to examine whether the frequent data users can be used as a reasonable training sample for model development. Fourth, in the current models, we assume all users share the same model parameters. However, different groups of people may have different mobility/telecommunication patterns and behavioral preferences. This may be accounted for in two steps---apply user clustering first and then develop models for each of the clusters. This requires a weaker assumption on the representativeness of training samples, as both the frequent data users and general population can be considered as different mixtures of the same underlying user clusters. More generally, a combination of unsupervised learning methods (e.g., \cite{chen_complete_2019}) and supervised learning methods (described in Section~\ref{sec:model}) can potentially further improve the model performance. Finally, in this case study, we only focus on inferring the existence of hidden visits. This is an important step of trip extraction from CDR data. An extension of this work is to infer, if a hidden visit exists, when and where it occurs. It is a more challenging problem. As each user visits a different set of locations at different time periods, the specific spatiotemporal patterns of hidden visits for one user may not be generalizable for other users. Future studies are needed to provide a better understanding on individual heterogeneity of spatiotemporal patterns. 

Whenever large-scale data is used, user privacy is an important consideration. The target application of our study is trip detection and OD estimation, which are done at aggregate level, not individual level. The developed models can be directly deployed on the database servers of telecom carriers, without need for data transfer. Furthermore, compared to other forms of big data, such as social media or credit card transaction data, CDR data is relatively less intrusive in terms of personal privacy. In addition, its localization error helps to mask the exact user locations, providing another layer of privacy preservation.

With the rapid advance in cellular network technologies and growth in smartphone usage, it is reasonable to expect that the data access records will be increasingly rich and prevalent. The specific parameters may change depending on the configuration of the networks or mobile phone settings, but the proposed data fusion approach is general and should still hold in the foreseeable future. Although CDR data are the focus of this study, the proposed approach can be extended to other types of coarse big data, such as transit smart card data and geo-located social media data. The value of this type of data can be enhanced through data fusion and statistical inference. For example, if we have smart card records and travel survey data for a group of individuals, the same data fusion approach may be applied to infer whether a user makes a hidden visit between two observed transit trips, which makes it possible to estimate individual-level mode shares for public transit.

\section*{Acknowledgements}
We would like to thank the Energy Foundation China and the China Sustainable Transportation Center for providing the financial support that made this research possible. 

\bibliographystyle{unsrt}  
\bibliography{CDR_Inference.bib}

\begin{thebibliography}{10}

\bibitem{mcnally_four-step_2007}
Michael~G. McNally.
\newblock The {Four}-{Step} {Model}.
\newblock In {\em Handbook of {Transport} {Modelling}}, volume~1 of {\em
  Handbooks in {Transport}}, pages 35--53. Emerald Group Publishing Limited,
  September 2007.

\bibitem{calabrese_urban_2014}
Francesco Calabrese, Laura Ferrari, and Vincent~D. Blondel.
\newblock Urban {Sensing} {Using} {Mobile} {Phone} {Network} {Data}: {A}
  {Survey} of {Research}.
\newblock {\em ACM Comput. Surv.}, 47(2):25:1--25:20, November 2014.

\bibitem{ranjan_are_2012}
Gyan Ranjan, Hui Zang, Zhi-Li Zhang, and Jean Bolot.
\newblock Are call detail records biased for sampling human mobility?
\newblock {\em ACM SIGMOBILE Mobile Computing and Communications Review},
  16(3):33, December 2012.

\bibitem{caceres_deriving_2007}
N.~Caceres, J.P. Wideberg, and F.G. Benitez.
\newblock Deriving origin destination data from a mobile phone network.
\newblock {\em IET Intelligent Transport Systems}, 1(1):15--26, March 2007.

\bibitem{calabrese_estimating_2011}
Francesco Calabrese, Giusy Di~Lorenzo, Liang Liu, and Carlo Ratti.
\newblock Estimating {Origin}-{Destination} {Flows} {Using} {Mobile} {Phone}
  {Location} {Data}.
\newblock {\em IEEE Pervasive Computing}, 10(4):36--44, April 2011.

\bibitem{mellegard_origin/destination-estimation_2011}
E.~Mellegard, S.~Moritz, and M.~Zahoor.
\newblock Origin/{Destination}-estimation {Using} {Cellular} {Network} {Data}.
\newblock In {\em 2011 {IEEE} 11th {International} {Conference} on {Data}
  {Mining} {Workshops} ({ICDMW})}, pages 891--896, December 2011.

\bibitem{wang_estimating_2013}
Ming-Heng Wang, Steven~D. Schrock, Nate~Vander Broek, and Thomas Mulinazzi.
\newblock Estimating {Dynamic} {Origin}-{Destination} {Data} and {Travel}
  {Demand} {Using} {Cell} {Phone} {Network} {Data}.
\newblock {\em International Journal of Intelligent Transportation Systems
  Research}, 11(2):76--86, April 2013.

\bibitem{iqbal_development_2014}
Md.~Shahadat Iqbal, Charisma~F. Choudhury, Pu~Wang, and Marta~C. Gonz{\'a}lez.
\newblock Development of origin{\textendash}destination matrices using mobile
  phone call data.
\newblock {\em Transportation Research Part C: Emerging Technologies},
  40:63--74, March 2014.

\bibitem{hasan_estimating_2017}
Md.~Mahedi Hasan and Mohammed~Eunus Ali.
\newblock Estimating {Travel} {Time} of {Dhaka} {City} from {Mobile} {Phone}
  {Call} {Detail} {Records}.
\newblock In {\em Proceedings of the {Ninth} {International} {Conference} on
  {Information} and {Communication} {Technologies} and {Development}}, {ICTD}
  '17, pages 1--11, New York, NY, USA, November 2017. Association for Computing
  Machinery.

\bibitem{ahas_using_2010}
Rein Ahas, Siiri Silm, Olle J{\"a}rv, Erki Saluveer, and Margus Tiru.
\newblock Using {Mobile} {Positioning} {Data} to {Model} {Locations}
  {Meaningful} to {Users} of {Mobile} {Phones}.
\newblock {\em Journal of Urban Technology}, 17(1):3--27, April 2010.

\bibitem{isaacman_identifying_2011}
Sibren Isaacman, Richard Becker, Ram{\'o}n C{\'a}ceres, Stephen Kobourov,
  Margaret Martonosi, James Rowland, and Alexander Varshavsky.
\newblock Identifying {Important} {Places} in {People}{\textquoteright}s
  {Lives} from {Cellular} {Network} {Data}.
\newblock In Kent Lyons, Jeffrey Hightower, and Elaine~M. Huang, editors, {\em
  Pervasive {Computing}}, number 6696 in Lecture {Notes} in {Computer}
  {Science}, pages 133--151. Springer Berlin Heidelberg, 2011.

\bibitem{gonzalez_understanding_2008}
Marta~C. Gonz{\'a}lez, C{\'e}sar~A. Hidalgo, and Albert-L{\'a}szl{\'o}
  Barab{\'a}si.
\newblock Understanding individual human mobility patterns.
\newblock {\em Nature}, 453(7196):779--782, June 2008.

\bibitem{phithakkitnukoon_activity-aware_2010}
Santi Phithakkitnukoon, Teerayut Horanont, Giusy~Di Lorenzo, Ryosuke Shibasaki,
  and Carlo Ratti.
\newblock Activity-{Aware} {Map}: {Identifying} {Human} {Daily} {Activity}
  {Pattern} {Using} {Mobile} {Phone} {Data}.
\newblock In Albert~Ali Salah, Theo Gevers, Nicu Sebe, and Alessandro
  Vinciarelli, editors, {\em Human {Behavior} {Understanding}}, number 6219 in
  Lecture {Notes} in {Computer} {Science}, pages 14--25. Springer Berlin
  Heidelberg, 2010.

\bibitem{schneider_unravelling_2013}
C.~M. Schneider, V.~Belik, T.~Couronne, Z.~Smoreda, and M.~C. Gonzalez.
\newblock Unravelling daily human mobility motifs.
\newblock {\em Journal of The Royal Society Interface},
  10(84):20130246--20130246, May 2013.

\bibitem{csaji_exploring_2013}
Bal{\'a}zs~Cs. Cs{\'a}ji, Arnaud Browet, V.A. Traag, Jean-Charles Delvenne,
  Etienne Huens, Paul Van~Dooren, Zbigniew Smoreda, and Vincent~D. Blondel.
\newblock Exploring the mobility of mobile phone users.
\newblock {\em Physica A: Statistical Mechanics and its Applications},
  392(6):1459--1473, March 2013.

\bibitem{zhao_explaining_2015}
Kai Zhao, Mirco Musolesi, Pan Hui, Weixiong Rao, and Sasu Tarkoma.
\newblock Explaining the power-law distribution of human mobility through
  transportation modality decomposition.
\newblock {\em Scientific Reports}, 5:9136, March 2015.

\bibitem{xu_uncovering_2018}
Yang Xu, Shih-Lung Shaw, Feng Lu, Jie Chen, and Qingquan Li.
\newblock Uncovering the {Relationships} {Between} {Phone} {Communication}
  {Activities} and {Spatiotemporal} {Distribution} of {Mobile} {Phone} {Users}.
\newblock In Shih-Lung Shaw and Daniel Sui, editors, {\em Human {Dynamics}
  {Research} in {Smart} and {Connected} {Communities}}, Human {Dynamics} in
  {Smart} {Cities}, pages 41--65. Springer International Publishing, Cham,
  2018.

\bibitem{zhao_individual-level_2016}
Zhan Zhao, Jinhua Zhao, and Haris~N. Koutsopoulos.
\newblock Individual-{Level} {Trip} {Detection} using {Sparse} {Call} {Detail}
  {Record} {Data} based on {Supervised} {Statistical} {Learning}.
\newblock 2016.
\newblock Number: 16-4386.

\bibitem{barabasi_origin_2005}
Albert-L{\'a}szl{\'o} Barab{\'a}si.
\newblock The origin of bursts and heavy tails in human dynamics.
\newblock {\em Nature}, 435(7039):207--211, May 2005.

\bibitem{bayir_mobility_2010}
Murat~Ali Bayir, Murat Demirbas, and Nathan Eagle.
\newblock Mobility profiler: {A} framework for discovering mobility profiles of
  cell phone users.
\newblock {\em Pervasive and Mobile Computing}, 6(4):435--454, August 2010.

\bibitem{chen_complete_2019}
Guangshuo Chen, Aline~Carneiro Viana, Marco Fiore, and Carlos Sarraute.
\newblock Complete trajectory reconstruction from sparse mobile phone data.
\newblock {\em EPJ Data Science}, 8(1):1--24, December 2019.
\newblock Number: 1 Publisher: SpringerOpen.

\bibitem{zhao_understanding_2016}
Ziliang Zhao, Shih-Lung Shaw, Yang Xu, Feng Lu, Jie Chen, and Ling Yin.
\newblock Understanding the bias of call detail records in human mobility
  research.
\newblock {\em International Journal of Geographical Information Science},
  30(9):1738--1762, September 2016.

\bibitem{jiang_review_2013}
Shan Jiang, Gaston~A. Fiore, Yingxiang Yang, Joseph Ferreira, Jr., Emilio
  Frazzoli, and Marta~C. Gonz{\'a}lez.
\newblock A {Review} of {Urban} {Computing} for {Mobile} {Phone} {Traces}:
  {Current} {Methods}, {Challenges} and {Opportunities}.
\newblock In {\em Proceedings of the {2Nd} {ACM} {SIGKDD} {International}
  {Workshop} on {Urban} {Computing}}, {UrbComp} '13, pages 2:1--2:9, New York,
  NY, USA, 2013. ACM.

\bibitem{hariharan_project_2004}
Ramaswamy Hariharan and Kentaro Toyama.
\newblock Project {Lachesis}: {Parsing} and {Modeling} {Location} {Histories}.
\newblock In Max~J. Egenhofer, Christian Freksa, and Harvey~J. Miller, editors,
  {\em Geographic {Information} {Science}}, number 3234 in Lecture {Notes} in
  {Computer} {Science}, pages 106--124. Springer Berlin Heidelberg, 2004.

\bibitem{chen_traces_2014}
Cynthia Chen, Ling Bian, and Jingtao Ma.
\newblock From traces to trajectories: {How} well can we guess activity
  locations from mobile phone traces?
\newblock {\em Transportation Research Part C: Emerging Technologies},
  46:326--337, September 2014.

\bibitem{chen_promises_2016}
Cynthia Chen, Jingtao Ma, Yusak Susilo, Yu~Liu, and Menglin Wang.
\newblock The promises of big data and small data for travel behavior (aka
  human mobility) analysis.
\newblock {\em Transportation Research Part C: Emerging Technologies},
  68:285--299, July 2016.

\bibitem{hoteit_filling_2016}
Sahar Hoteit, Guangshuo Chen, Aline Viana, and Marco Fiore.
\newblock Filling the gaps: on the completion of sparse call detail records for
  mobility analysis.
\newblock In {\em Proceedings of the {Eleventh} {ACM} {Workshop} on
  {Challenged} {Networks}}, {CHANTS} '16, pages 45--50, New York, NY, USA,
  October 2016. Association for Computing Machinery.

\bibitem{chen_enriching_2018}
Guangshuo Chen, Sahar Hoteit, Aline~Carneiro Viana, Marco Fiore, and Carlos
  Sarraute.
\newblock Enriching sparse mobility information in {Call} {Detail} {Records}.
\newblock {\em Computer Communications}, 122:44--58, June 2018.

\bibitem{james_introduction_2013}
Gareth James, Daniela Witten, Trevor Hastie, and Robert Tibshirani.
\newblock {\em An {Introduction} to {Statistical} {Learning}}, volume 103 of
  {\em Springer {Texts} in {Statistics}}.
\newblock Springer New York, New York, NY, 2013.

\bibitem{eagle_reality_2006}
Nathan Eagle and Alex~(Sandy) Pentland.
\newblock Reality mining: sensing complex social systems.
\newblock {\em Personal and Ubiquitous Computing}, 10(4):255--268, May 2006.

\bibitem{williams_measures_2015}
Nathalie~E. Williams, Timothy~A. Thomas, Matthew Dunbar, Nathan Eagle, and
  Adrian Dobra.
\newblock Measures of {Human} {Mobility} {Using} {Mobile} {Phone} {Records}
  {Enhanced} with {GIS} {Data}.
\newblock {\em PLOS ONE}, 10(7):e0133630, July 2015.

\bibitem{candia_uncovering_2008}
J.~Candia, M.~C. Gonz{\'a}lez, P.~Wang, T.~Schoenharl, G.~Madey, and A.-L.
  Barab{\'a}si.
\newblock Uncovering individual and collective human dynamics from mobile phone
  records.
\newblock {\em Journal of Physics A: Mathematical and Theoretical},
  41(22):224015, June 2008.
\newblock arXiv: 0710.2939.

\bibitem{alexander_origindestination_2015}
Lauren Alexander, Shan Jiang, Mikel Murga, and Marta~C. Gonz{\'a}lez.
\newblock Origin{\textendash}destination trips by purpose and time of day
  inferred from mobile phone data.
\newblock {\em Transportation Research Part C: Emerging Technologies}, 58, Part
  B:240--250, September 2015.

\bibitem{friedman_greedy_2001}
Jerome~H. Friedman.
\newblock Greedy function approximation: {A} gradient boosting machine.
\newblock {\em Annals of Statistics}, 29(5):1189--1232, October 2001.
\newblock Publisher: Institute of Mathematical Statistics.

\bibitem{pedregosa_scikit-learn:_2011}
Fabian Pedregosa, Ga{\"e}l Varoquaux, Alexandre Gramfort, Vincent Michel,
  Bertrand Thirion, Olivier Grisel, Mathieu Blondel, Peter Prettenhofer, Ron
  Weiss, Vincent Dubourg, Jake Vanderplas, Alexandre Passos, David Cournapeau,
  Matthieu Brucher, Matthieu Perrot, and {\'E}douard Duchesnay.
\newblock Scikit-learn: {Machine} {Learning} in {Python}.
\newblock {\em Journal of Machine Learning Research}, 12:2825--2830, October
  2011.

\bibitem{song_limits_2010}
Chaoming Song, Zehui Qu, Nicholas Blumm, and Albert-L{\'a}szl{\'o}
  Barab{\'a}si.
\newblock Limits of {Predictability} in {Human} {Mobility}.
\newblock {\em Science}, 327(5968):1018--1021, February 2010.

\bibitem{platt_probabilistic_1999}
John~C. Platt.
\newblock Probabilistic {Outputs} for {Support} {Vector} {Machines} and
  {Comparisons} to {Regularized} {Likelihood} {Methods}.
\newblock In {\em {ADVANCES} {IN} {LARGE} {MARGIN} {CLASSIFIERS}}, pages
  61--74. MIT Press, 1999.

\end{thebibliography}

\end{document}